\documentclass{article}
\usepackage{spconf,amsmath,graphicx}
\usepackage{amsfonts,amssymb}
\usepackage{bm}
\usepackage{subfigure}
\usepackage{float}
\usepackage{tikz,pgfplots}
\usepackage{bm}
\usepackage{cases}
\usepackage{algorithm}
\usepackage{algorithmic}
\usepackage{multirow}
\usepackage{cite,bbm,booktabs,url}
\usepackage{cases}
\usepackage{theorem}
\usepackage{enumitem}

\usepackage[colorlinks=true,
linkcolor=blue,
urlcolor=blue,
citecolor=blue]{hyperref}  

\usepackage{cleveref}


\usepackage{shortcuts_OPT}

\newtheorem{Lemma}{Lemma}

\newtheorem{Def}{Definition}
\newtheorem{Conjecture}{Conjecture}

\theorembodyfont{\rmfamily}

\newtheorem{assumption}{H\!\!}

\crefname{Def}{Definition}{Definition} 

\title{Identifying First-order Lowpass Graph Signals using Perron Frobenius Theorem}
\name{Yiran He, Hoi-To Wai\thanks{Emails: \texttt{\{yrhe,htwai\}@se.cuhk.edu.hk}}}
\address{Department of SEEM, The Chinese University of Hong Kong, Shatin, Hong Kong SAR of China}
\begin{document}
\ninept
\maketitle
\begin{abstract}
This paper is concerned with the blind identification of graph filters from graph signals.
Our aim is to determine if the graph filter generating the graph signals is \emph{first-order lowpass} without knowing the graph topology. 
Notice that lowpass graph filter is a common pre-requisite for applying graph signal processing tools for sampling, denoising, and graph learning. 
Our method is inspired by the Perron Frobenius theorem, which observes that for first-order lowpass graph filter, the top eigenvector of output covariance would be the \emph{only eigenvector} with elements of the same sign. 
Utilizing this observation, we develop a simple detector that answers if a given data set is produced by a first-order lowpass graph filter. We analyze the effects of finite-sample, graph size, observation noise, strength of lowpass filter, on the detector's performance. Numerical experiments on synthetic and real data support our findings. 
\end{abstract}


\section{Introduction} 
A recent trend in data science is to develop tools for analyzing and inference from signals defined on graph, a.k.a.~graph signals. Examples are financial and social networks data where graph structures can be leveraged to improve inference  \cite{kolaczyk2014statistical,shuman2013emerging}. 
These premises motivated researches on graph signal processing (GSP) \cite{Sand2013DSP,ortega2018graph} which extends signal processing models to graph signals.

To study graph signals, an important concept of GSP is to model them as outputs from exciting a graph filter. The graph filter captures the social/physical process that generates the observations we collect. Examples are heat diffusion \cite{thanou2017learning}, dynamics of functional brain activities \cite{huang2018graph}, equilibrium seeking in network games \cite{wai2019blind}, etc.. Similar to its linear time invariant counterpart, the graph filters can be classified as \emph{lowpass}, \emph{bandpass}, \emph{highpass} according to their frequency responses computed through the Graph Fourier transform \cite{sandryhaila2014discrete}.
Among others, lowpass graph filters and signals are important to many GSP tools, e.g., sampling, denoising, graph topology learning \cite{ramakrishna2020user}. Importantly, a number of practical social/physical models can naturally lead to lowpass graph signals as studied in \cite{ramakrishna2020user}. 
% it is a reasonable model as argued in \cite{ramakrishna2020user}, where the lowpass property is shared by a number of social/physical models. 

Although lowpass graph filters/signals are common, most prior works took the lowpass property as a default assumption without further verification.
This paper takes a different approach, where we address the validity of the lowpass assumption for a given set of graph signals. 
Our approach is \emph{data-driven}, and 
% to test whether a set of graph signals are generated from a lowpass graph filter.
we aim to provide certificates on whether certain GSP tools are applicable to a dataset whose social/physical models are unknown, as well as determining the {type} of social/physical process which supports the observed data. 
% Specifically, we identify graph signals from \emph{first-order lowpass graph filters} which attenuate every graph frequencies above the lowest one. 

Our work is related to the recent developments in \emph{graph inference or learning} which aim at learning the graph topology blindly \cite{kalofolias2016learn,dong2016learning,egilmez2017graph,pasdeloup2017characterization,wai2019blind,Wai2020BCest,Segarra2020BCest}. For instance, \cite{kalofolias2016learn,dong2016learning,egilmez2017graph,pasdeloup2017characterization} infer topology from smooth graph signals which can be given by lowpass graph filtering, \cite{wai2019blind,Wai2020BCest,Segarra2020BCest} consider blind inference of communities and centrality from lowpass graph signals; see \cite{mateos2019connecting,dong2019learning}.
However, these works rely on the lowpass property as \emph{an assumption} on the graph signals, while the aim of this paper is to verify the latter property.
Recent works also considered the joint inference of graph topology and network dynamics \cite{ioannidis2019semi,wai2019joint}, inferring the type of physical process \cite{barzel2013universality}, or the identification of graph filters in \cite{segarra2016blind,zhu2020estimating}. These works either require the graph topology is known a priori, or the network dynamics is parameterized with a specific model template. 

% segarra2016blind needs the graph !!

% related works includes

% the setup is motivated by our works (like centrality, community, etc..)!

Our contributions are three-fold. First, in Section~\ref{sec:model}, we demonstrate that for first-order lowpass graph signals, the Perron Frobenius theorem shows that the top eigenvector of the covariance matrix must have elements of the same sign. Furthermore, this is the only such eigenvector. Second, in Section~\ref{sec:det}, we design a simple detector for lowpass graph signals and analyze the effects of finite-sample on its performance. Importantly, we show that for strong lowpass graph filters, the detection performance is robust to the graph size and/or number of samples. 
Third, in Section~\ref{sec:num}, we apply our detector on three real dataset (S\&P500 stock, US Senate's voting record, number of new COVID-19 cases) and confirm if the underlying graph filter is lowpass. 
Our work provides the first step towards blind identification of network dynamics without knowing graph topology.\vspace{-.1cm}

\section{Graph Signals and Graph Filters}\vspace{-.1cm} \label{sec:model}
This section describes the graph signal model and introduces necessary notations.
We consider an undirected graph denoted by $G = (V,E)$, with $n$ nodes described in $V = \{1,...,n\}$ and $E \subseteq V \times V$ is the edge set. We assume $(i,i) \notin E$ such that $G$ has no self-loop. Define the weighted adjacency matrix ${\bm A} \in \Re^{n \times n}$ as a non-negative, symmetric matrix with $A_{ij} > 0$ if and only if $(i,j) \in E$. The Laplacian matrix is defined as ${\bm L} = {\rm Diag}( {\bm A} {\bf 1} ) - {\bm A}$. 

A graph signal \cite{Sand2013DSP} on $G$ is a scalar function defined on $V$, i.e., $x : V \rightarrow \RR$, and it can be represented as an $n$-dimensional vector ${\bm x} \in \RR^n$. We use graph filters to describe an unknown process on $G$, capturing phenomena such as information exchange, diffusion, etc.. A linear graph filter can be expressed as a $T$th order matrix polynomial:\vspace{-.1cm}
\beq \textstyle
{\cal H}( {\bm S} ) = \sum_{t=0}^{T-1} h_t {\bm S}^t,\vspace{-.0cm}
\eeq
where $\{ h_t \}_{t=0}^{T-1}$ is the set of filter weights and ${\bm S}$ is the graph shift operator (GSO) which is a symmetric matrix that respects the graph topology. In this paper, the GSO ${\bm S}$ can be either adjacency matrix ${\bm A}$, or Laplacian matrix ${\bm L}$. In both cases, we consider its eigendecomposition as ${\bm S} = {\bm V} \bm{\Lambda} {\bm V}^\top$, where ${\bm V}$ is orthogonal and $\bm{\Lambda} = {\rm Diag}(\lambda_1,...,\lambda_n)$ is a diagonal matrix of the eigenvalues of ${\bm S}$. 
The $i$th column of ${\bm V}$ satisfies ${\bm S} {\bm v}_i = \lambda_i {\bm v}_i$.
Define the frequency response function:\vspace{-.1cm}
\beq \textstyle \label{eq:freq}
h(\lambda) := \sum_{t=0}^{T-1} h_t \lambda^t.\vspace{-.0cm}
\eeq
Alternatively, the graph filter can be written in terms of its frequency response as ${\cal H}( {\bm S}) = {\bm V} h(\bm{\Lambda}) {\bm V}^\top$, where $h(\bm{\Lambda})$ is a diagonal matrix with $[h(\bm{\Lambda})]_{ii} = h(\lambda_i)$. 

In GSP, a \emph{low frequency} graph signal is one that varies little along the edges on $G$, i.e., with a small graph $\ell_2$ total variation (graph TV) ${\bm x}^\top {\bm L} {\bm x}$ \cite{ortega2018graph}. For Laplacian matrix ${\bm L}$, it is known that for small eigenvalue $\lambda_i$, the corresponding eigenvector ${\bm v}_i$ is a low frequency graph signal, e.g., ${\bm L} (1/\sqrt{n} {\bm 1}) = 0 \cdot (1/\sqrt{n} {\bm 1})$.
This observation will be reversed for adjacency matrix ${\bm A}$, where the eigenvector with a large $\lambda_i$ corresponds to a low frequency graph signal. 

With a slight abuse of notations, we adopt the convention that when ${\bm S} = {\bm L}$, the eigenvalues (a.k.a.~graph frequencies) are ordered as $0 = \lambda_1 \leq \ldots \leq \lambda_n$; while when ${\bm S} = {\bm A}$, the eigenvalues are ordered as $\lambda_1 \geq \ldots \geq \lambda_n$.
We now define the lowpass graph filter for a general GSO ${\bm S}$ \cite{ramakrishna2020user} as follows:\vspace{-.2cm}
\begin{Def} \label{def:lpf}
Let $K \in \{1,...,n\}$. A graph filter ${\cal H}({\bm S})$ is lowpass with cutoff frequency at $\lambda_K$ and the lowpass ratio $\eta_K < 1$ if\vspace{-.1cm}
\beq \notag
\max\{ | h(\lambda_{K+1}) |, ..., | h(\lambda_n) | \} =  \eta_K  \min\{ | h(\lambda_1) |, ..., |h(\lambda_K)| \},\vspace{-.1cm}
\eeq
where the frequency response function $h(\cdot)$ was defined in \eqref{eq:freq}. \vspace{-.1cm}
\end{Def}
% Without loss of generality, we assume $h(\lambda_1) > 0$.
The lowpass ratio $\eta_K$ determines the strength of the lowpass filter as it quantifies the degree of attenuation beyond cutoff frequency. We say ${\cal H}({\bm S})$ is strong (resp.~weak) lowpass if $\eta_K \ll 1$ (resp.~$\eta_K \approx 1$).
 
Our task is to identify whether a set of graph signals was generated from a lowpass graph filter, i.e., if they are \emph{lowpass graph signals}. The observed graph signals are modeled as the noisy {filter outputs} from ${\cal H}( {\bm S} )$ subjected to excitations $\{ {\bm x}_\ell \}_{\ell=1}^m$. We have:\vspace{-.1cm}
\beq \label{eq:graphsig}
{\bm y}_\ell = \overline{\bm y}_\ell + {\bm w}_\ell,~\text{where}~~\overline{\bm y}_\ell = {\cal H}( {\bm S} ) {\bm x}_\ell,~\ell=1,...,m,\vspace{-.1cm}
\eeq
such that ${\bm w}_\ell$ is the observation noise, and ${\bm x}_\ell, {\bm w}_\ell$ are zero mean random vectors with the covariances $\EE[ {\bm x}_\ell {\bm x}_\ell^\top ] ={\bm I}$, $\EE[ {\bm w}_\ell {\bm w}_\ell^\top ] = \sigma^2 {\bm I}$.
In the above, the signal term $\overline{\bm y}_\ell = \sum_{t=0}^{T-1} h_t {\bm S}^t {\bm x}_\ell$ is a weighted sum of the shifted versions of ${\bm x}_\ell$, i.e., it is the output of the graph filter under the excitation ${\bm x}_\ell$. 
Note that we consider a \emph{completely blind identification} problem where it is not known a priori if the GSO is a Laplacian matrix or an adjacency matrix.

\subsection{Perron Frobenius Theorem and 1st-Order Lowpass Filter}
While the general problem is to identify lowpass graph filters of \emph{any order}, this paper focuses on the \emph{first-order lowpass graph filters} whose cutoff frequency is $\lambda_1$ with the lowpass ratio of $\eta_1 < 1$ [cf.~\Cref{def:lpf}].
Notice that this is a sufficient condition to enable graph inference such as blind centrality estimation \cite{Wai2020BCest, Segarra2020BCest}, as well as ensuring that the graph signal is smooth \cite{dong2016learning,kalofolias2016learn}. 
Formally, our task is to distinguish between the following two hypothesis:\vspace{-.1cm}
\beq
\begin{split}
{\cal T}_0 & : \text{the graph filter ${\cal H}({\bm S})$ is first-order lowpass} \\
{\cal T}_1 & : \text{the graph filter ${\cal H}({\bm S})$ is \emph{not} first-order lowpass}\\[-.1cm]
\end{split}
\eeq
Note that ${\cal T}_1$ include highpass filters where \Cref{def:lpf} is not satisfied for any $K \in \{1,...,n\}$, as well as lowpass filters with a higher cutoff frequency at $\lambda_K$, $K \geq 2$.

Our next endeavor is to study the spectral property of the covariance of $\{ {\bm y}_\ell \}_{\ell=1}^m$. For simplicity, let us focus on the signal term $\overline{\bm y}_\ell$ with the following population covariance matrix:\vspace{-.1cm}
\beq \label{eq:cov}
{\rm Cov}( \overline{\bm y}_\ell ) = \EE[ \overline{\bm y}_\ell (\overline{\bm y}_\ell)^\top ] = {\cal H}({\bm S})^2 = {\bm V} h( \bm{\Lambda} )^2 {\bm V}^\top.\vspace{-.1cm}
\eeq
Let $\overline{\bm v}_i$ be the top $i$th eigenvector of ${\rm Cov}( \overline{\bm y}_\ell )$ with $\| \overline{\bm v}_i \| = 1$. 
Under the assumption that ${\cal H}({\bm S})$ is first-order lowpass [cf.~\Cref{def:lpf}], it is obvious that we have $\overline{\bm v}_1 = {\bm v}_1$, i.e., the lowest frequency graph signal that varies little along the edges on $G$.
The above suggests that one could detect if ${\cal H}({\bm S})$ is a first-order lowpass graph filter by evaluating the graph TV of $\overline{\bm v}_1$ (or its approximation computed from $\{ {\bm y}_\ell \}_{\ell=1}^m$) and applying a threshold detector. However, it is not possible to evaluate the graph TV $\overline{\bm v}_1^\top {\bm L} \overline{\bm v}_1$ since ${\bm L}$ is unknown.  
% 1st order LPF is an approximation of LPF?!

Instead of computing the graph TV of $\overline{\bm v}_1$, we inspect the eigenvector ${\bm v}_1$ of the lowest frequency of the GSO. Assume that:\vspace{-.3cm}
\begin{assumption} \label{ass:connected}
The graph $G$ has one and only one connected component.\vspace{-.1cm} 
\end{assumption}
\Cref{ass:connected} is common and it allows us to characterize the lowest graph frequency eigenvector of ${\bm L}$, ${\bm A}$.
Since ${\bm A}$ is non-negative, and ${\bm L}$ has non-positive off-diagonal elements, applying the Perron-Frobenius theorem shows that the lowest graph frequency eigenvectors must be positive \cite{horn2012matrix}. In particular, we observe:\vspace{-.1cm}
\begin{Lemma} \label{lem:gso} Under \Cref{ass:connected}, it holds that:
\begin{enumerate}[leftmargin=4mm]
\item For Laplacian matrix ${\bm L}$, the (smallest) eigenvalue $\lambda_1 = 0$ has multiplicity one with the eigenvector ${\bm v}_1 = {\bf 1} / \sqrt{n}$. For adjacency matrix ${\bm A}$, the (largest) eigenvalue $\lambda_1$ has multiplicity one with the eigenvector ${\bm v}_1$, which is a positive vector.
\item In addition, ${\bm v}_1$ is the only positive eigenvector\footnote{Note that both ${\bm v}_1, -{\bm v}_1$ are eigenvectors with the eigenvalue $\lambda_1$. We assume ${\bm v}_1 > {\bm 0}$ to avoid such ambiguity.} of ${\bm L}$ or ${\bm A}$. For $i \neq 1$, the eigenvector ${\bm v}_i \neq {\bm 0}$ must have  at least one positive and one negative element.
\end{enumerate}
\end{Lemma}
\emph{Proof}: 
The first statement is a consequences of \Cref{ass:connected} and the spectral graph theory for ${\bm L}$ \cite[Theorem 7.1.2]{cvetkovic2009introduction}, or the Perron-Frobenius theorem for ${\bm A}$ \cite[Theorem 8.4.4]{horn2012matrix}. Notice that the positivity of ${\bm v}_1$ for ${\bm L}$ can alternatively be shown by the Perron-Frobenius theorem through studying $\upsilon {\bm I} - {\bm L}$ with sufficiently large $\upsilon>0$.

To show the second statement, we use the orthogonality of ${\bm V}$ which implies ${\bm v}_i^\top {\bm v}_1 = 0$ for any $i \neq 1$. For the sake of contradiction, assume ${\bm v}_i \neq {\bm 0}$, ${\bm v}_i \geq {\bm 0}$ (resp.~${\bm v}_i \leq {\bm 0}$). Since ${\bm v}_1 > {\bm 0}$, we must have ${\bm v}_i^\top {\bm v}_1 > 0$ (resp.~${\bm v}_i^\top {\bm v}_1 < 0$), leading to a contradiction.\hfill$\square$\vspace{.2cm}

Consider the \emph{null hypothesis} ${\cal T}_0$, 
using \Cref{lem:gso} and \eqref{eq:cov}, we observe the top eigenvector $\overline{\bm v}_1$ of ${\rm Cov}( {\bm y}_\ell )$ must be a \emph{positive vector}.
For the \emph{alternative hypothesis} ${\cal T}_1$, we observe from \eqref{eq:cov} that $\overline{\bm v}_1$ will be taken as one of the other eigenvectors, ${\bm v}_i$, of ${\bm S}$ with $i \neq 1$. By \Cref{lem:gso}, $\overline{\bm v}_1$ must have at least one negative and positive element.\vspace{-.1cm}

% The above discussions highlight the spectral feature of ${\rm Cov}( \overline{\bm y}_\ell )$ under different hypothesis. The next section will focus on designing a detector that distinguishes between the two hypothesis on the graph filter ${\cal H}( {\bm S})$.

% Below is a summary;
% \begin{Observation}
% Under \Cref{ass:connected}. Let $\overline{\bm v}_1$ be the top eigenvector of ${\rm Cov}({\bm y}_\ell)$. 
% If the graph filter ${\cal H}({\bm S})$ is 1st-order lowpass, then the top eigenvector must be strictly positive (or negative) $\overline{\bm v}_1 > 0$; otherwise, it must contain a negative element $\overline{\bm v}_1 \ngtr 0$.
% \end{Observation}

% I will rewrite the last paragraph here...

\section{Identifying Low-pass Graph Signals}\vspace{-.1cm} \label{sec:det}
In this section we propose heuristics to detect first-order lowpass graph signals and provide insights into its performance. 

The discussions from the previous section suggest that one could distinguish between ${\cal T}_0, {\cal T}_1$ by inspecting whether all elements of the top eigenvector of ${\rm Cov}({\bm y}_\ell)$ have the same sign. We define:\vspace{-.1cm}
\beq \label{eq:samp_cov} \textstyle
\widehat{\bm C}_y^m \eqdef (1/m) \sum_{\ell=1}^m {\bm y}_\ell ({\bm y}_\ell)^\top, ~~ \widehat{\bm v}_i = \textsf{$i$th-EV}( \widehat{\bm C}_y^m ),\vspace{-.1cm}
\eeq
such that $\widehat{\bm C}_y^m$ is the sampled covariance and $\widehat{\bm v}_i$ is the latter's eigenvector with the $i$th largest eigenvalue.
We further define the scoring function:\vspace{-.2cm}
\beq \label{eq:stat}
\Gamma( {\bm v} ) \eqdef \min \big\{ \| {\bm v} - ({\bm v} )_+ \|_2, \| {\bm v} + (-{\bm v})_+ \} \|_2  \big\} \vspace{-.1cm}
\eeq
where $({\bm v})_+ = \max\{ {\bm v}, {\bm 0} \}$ is the elementwise maximum of ${\bm v}$ and ${\bm 0}$.
In both cases, we observe that ${\bm v} = ({\bm v} )_+$ if ${\bm v}$ is positive, and ${\bm v} = - (-{\bm v} )_+$ if ${\bm v}$ is negative. As such, $\Gamma( {\bm v} ) = 0$ if and only if ${\bm v}$ is a positive or negative vector; otherwise, $\Gamma( {\bm v} ) > 0$. 
% We remark it is also natural to adopt alternative norms such as $\ell_\infty$ norms in \eqref{eq:stat}.
% They will be examined numerically in the next section.

Based on the scoring function \eqref{eq:stat}, we propose the following heuristic to detect if a set of graph signals are first-order lowpass filtered. Let $\widehat{\cal T}$ be the detector output, we have:\vspace{-.1cm}
\beq \label{eq:detector}
\widehat{\cal T} = \begin{cases}
{\cal T}_0 &,~\text{if}~\Gamma( \widehat{\bm v}_1 ) \leq \min_{i=2,...,n} \Gamma( \widehat{\bm v}_i ), \\
{\cal T}_1 &,~\text{otherwise}.
\end{cases}\vspace{-.1cm}
\eeq
Note that the detector is inspired by the observation in \Cref{lem:gso} that under ${\cal T}_0$, the top eigenvector of the signal term's population covariance must be positive and it is the only such eigenvector.\vspace{-.2cm}

\subsection{Insights from Performance Analysis}
Analyzing the performance of the detector \eqref{eq:detector} is challenging as it involves the order statistics of $\{ \Gamma(\widehat{\bm v}_i) \}_{i=1}^n$. 
We instead provide insights on the performance of \eqref{eq:detector} through analyzing the effects of finite-sample, graph size and noise variance.

We assume $\widehat{\bm v}_j^\top \overline{\bm v}_j \geq 0$ without loss of generality. 
We adopt the Davis-Kahan theorem from \cite[Corollary 3]{yu2015useful} as follows:\vspace{-.2cm}
\begin{Lemma} \label{lem:dk} 
Let $j \in \{1,...,n\}$, consider the eigenvectors $\overline{\bm v}_j$, $\widehat{\bm v}_j$ from the population, sampled covariance. Let $\widehat{\bm v}_j^\top \overline{\bm v}_j \geq 0$, it holds \vspace{-.1cm}
\beq \label{eq:bd}
\| \widehat{\bm v}_j - \overline{\bm v}_j \|_2 \leq \frac{ 2^{3/2} \| \widehat{\bm C}_y^m - {\rm Cov}( \overline{\bm y}_\ell ) \|_2 }{ \min\{ \beta_{j-1} - \beta_j, \beta_j - \beta_{j+1} ) ) } \vspace{-.1cm}
\eeq
where $\beta_j$ denotes the $j$th largest eigenvalue of the population covariance ${\rm Cov}( \overline{\bm y}_\ell )$ with the convention $\beta_0 = \infty, \beta_{n+1} = -\infty$.\vspace{-.1cm}
\end{Lemma}
Let $r = {\rm Tr}({\rm Cov}( \overline{\bm y}_\ell )) / \| {\rm Cov}( \overline{\bm y}_\ell ) \|_2$, the denominator in the r.h.s.~of \eqref{eq:bd} can be bounded with high probability \cite[Remark 5.6.3]{vershynin2018high}: \vspace{-.1cm}
\beq \label{eq:sample_bd}
\begin{split}
\| \widehat{\bm C}_y^m - {\rm Cov}( \overline{\bm y}_\ell ) \|_2
& \textstyle \leq \sigma^2 + \| \widehat{\bm C}_y^m - (1/m)\sum_{\ell=1}^m {\bm y}_\ell {\bm y}_\ell^\top \|_2 \\
& = \sigma^2 + {\cal O}( \| {\rm Cov}( \overline{\bm y}_\ell ) \|_2 \sqrt{r/m}). \\[-.1cm]
\end{split}
\eeq
Note that $r$ is the effective rank of ${\rm Cov}( \overline{\bm y}_\ell )$ which is close to 1 for strong lowpass filters with $\eta_1 \ll 1$, yet for weak lowpass filter with $\eta_1 \approx 1$, one has $r \approx n$. 
In both situations, this shows $\widehat{\bm v}_j \approx \overline{\bm v}_j$ with small noise and large number of samples. 

For $j \neq 1$, it is known that $\| {\bm v}_j - ( {\bm v}_j )_+ \|_2 > 0$ where ${\bm v}_j$ is the $j$th eigenvector of ${\bm S}$. However, it is not clear how large should this value be.  
To this end, we state the following conjecture:\vspace{-.2cm}
\begin{Conjecture} \label{conj:hpf}
For $j \neq 1$, we have $\| {\bm v}_j - ( {\bm v}_j )_+ \|_2 = \Theta(1)$.\vspace{-.1cm}
\end{Conjecture}
The conjecture states that the magnitude of $\| {\bm v}_j - ( {\bm v}_j )_+ \|_2$ is independent of the graph size $n$. Our rationale is that the vector ${\bm v}_j$ is `non-localized' whose energy is evenly spread and there are ${\cal O}(n)$ negative elements, e.g., see \cite{fiedler1975property} for insights behind the conjecture.
% With the above observations, we will show that the performance of \eqref{eq:detector} improves with {\sf (i)} the lowpass ratio $\eta_1$, {\sf (ii)} number of samples $m$ and noise variance $\sigma^2$, and {\sf (iii)} graph size $n$.
\vspace{.2cm}

\noindent \textbf{Case ${\cal T}_0$}. We consider the null hypothesis when ${\cal H}({\bm S})$ is a first-order lowpass filer. Observe $\overline{\bm v}_1 > {\bm 0}$ and we have\vspace{-.1cm}
\beq \notag
\begin{split}
\Gamma( \widehat{\bm v}_1 ) & = \| \widehat{\bm v}_1 - (\widehat{\bm v}_1)_+ \|_2 \leq 2 \| \widehat{\bm v}_1 - \overline{\bm v}_1 \|_2 + \| \overline{\bm v}_1 - ( \overline{\bm v}_1 )_+ \|_2 \\[-.1cm]
\end{split}
\eeq
where we have applied $\| ({\bm v})_+ - ({\bm v}')_+ \|_2 \leq \| {\bm v} - {\bm v}' \|_2 $ in the first inequality. Furthermore, we have \vspace{-.1cm}
\beq \notag
\begin{split}
\beta_1 - \beta_2 & \textstyle = | h( \lambda_1 ) |^2 - \max_{i=2,...,n}| h(\lambda_i) |^2 = | h( \lambda_1 ) |^2 (1 - \eta_1).
\end{split}
\eeq
Combining \Cref{lem:dk} and \eqref{eq:sample_bd} yields the upper bound: 
\beq \label{eq:gamma1}
\Gamma( \widehat{\bm v}_1 ) \leq 2^{5/2} (1-\eta_1)^{-1} \big\{ \sigma^2 + {\cal O}(\sqrt{r/m}) \big\}.
\eeq
where we used $\| {\rm Cov}( \overline{\bm y}_\ell ) \|_2 = |h(\lambda_1)|^2$ to simplify the expression.

On the other hand, for any $j \neq 1$, we have the following lower bound to the scoring function $\Gamma(\widehat{\bm v}_j)$:
\beq \label{eq:gammaj}
\begin{split}
\Gamma( \widehat{\bm v}_j ) & = \| \widehat{\bm v}_j - ( \widehat{\bm v}_j )_+ \|_2 \geq \| \overline{\bm v}_j - ( \overline{\bm v}_j)_+ \|_2 - 2 \| \widehat{\bm v}_j - \overline{\bm v}_j \|_2 \\
& \geq \Theta(1) - \frac{2^{5/2} ( \sigma^2 + {\cal O}(|h(\lambda_1)|^2\sqrt{r/m}) ) }{\min\{ \beta_{j-1}-\beta_j, \beta_j - \beta_{j+1} \}}.
\end{split}
\eeq
where the last inequality is due to \Cref{conj:hpf} and \Cref{lem:dk}. 
\vspace{.2cm}

\noindent \textbf{Case ${\cal T}_1$}. For the alternative hypothesis where ${\cal H}({\bm S})$ is \emph{not a first-order lowpass filter}. 
We observe that $\overline{\bm v}_1$ is no longer equal to the lowest frequency eigenvector of ${\bm S}$, i.e., ${\bm v}_1$. Similar to \eqref{eq:gammaj}, this yields the following lower bound of $\Gamma( \widehat{\bm v}_1 )$:
\beq \label{eq:gamma1_t1}
\Gamma( \widehat{\bm v}_1 ) \geq \Theta(1) - 2^{5/2} (\beta_1 - \beta_2)^{-1} (\sigma^2+ {\cal O}(|h(\lambda_1)|^2\sqrt{r/m})) 
\eeq
On the other hand, there exists $j\neq 1$,  $\overline{\bm v}_j = {\bm v}_1$. Similar to \eqref{eq:gamma1}, we observe\vspace{-.2cm}
\beq \label{eq:gammaj_t1}
\Gamma( \widehat{\bm v}_j ) \leq \frac{2^{5/2} ( \sigma^2 + {\cal O}(|h(\lambda_1)|^2\sqrt{r/m}) )}{ \min\{ \beta_{j-1} -\beta_j , \beta_j - \beta_{j+1} \} }.
\eeq
% the effect of eta_1 

% \noindent \textbf{Discussions}.
Comparing the observations in \eqref{eq:gamma1}--\eqref{eq:gammaj_t1} shows that the detector \eqref{eq:detector} has a low error rate when {\sf (i)} the observation noise $\sigma$ is small, {\sf (ii)} the number of samples $m$ is large, which are the expected behaviors.
On the other hand, the effects of graph size $n$ is not immediately clear. We observe from \eqref{eq:gamma1} that the detection performance is insensitive to $n,m, \sigma^2$ with a strong lowpass filter that has $\eta_1 \ll 1$ (and thus $r \approx 1$). On the other hand, the detection performance may degrade with a weak lowpass filter since $\eta_1 \approx 1$ (and thus $r \approx n$). \vspace{-.1cm}

\begin{figure}[t]
\centering
\resizebox{.52\linewidth}{!}{% This file was created by tikzplotlib v0.8.2.
\begin{tikzpicture}

\definecolor{color0}{rgb}{0,0.75,0.75}
\definecolor{color1}{rgb}{0.75,0.75,0}

\begin{axis}[
legend cell align={left},
legend columns=1,
legend style={fill opacity=0.8, draw opacity=1, text opacity=1,at={(0.8,1.0)}, anchor=north, draw=white!80.0!black},
log basis x={10},
log basis y={10},
tick align=outside,
tick pos=both,
title={},
x grid style={white!69.01960784313725!black},
xlabel={\large Sample size $m$},
xmajorgrids,
xmin=7.67270499010925, xmax=2606.64264112612,
xmode=log,
xtick style={color=black},
y grid style={white!69.01960784313725!black},
ylabel={\large Scoring fct.~$\Gamma(\widehat{\bm v}_1)$},
ymajorgrids,
ymin=-0.0134200012543306, ymax=0.8,
%ymode=log,
ytick style={color=black},
width=7.5cm,height=4cm,
grid=both,grid style={line width=.2pt, draw=gray!50},
]
\addplot [thick, blue, mark=o, mark size=3, mark options={solid}]
table {%
10 0.5976714541872754
50 0.5422115725800624
100 0.40990978041580595
500 0.07827005194817047
1000 0.02005137707345903
2000 0.0035296529736202715
};
\addlegendentry{Weak L. Lap.}
\addplot [thick, red, mark=o, mark size=3, mark options={solid}]
table {%
10 0.62263730327884
50 0.6360137262243512
100 0.581625438909376
500 0.29294624675345565
1000 0.1424681072709853
2000 0.03563287484762166
};
\addlegendentry{Weak L. Adj.}
%\addplot [thick, black, mark=o, mark size=3, mark options={solid}]
%table {%
%10 0.22778566913983717
%50 0.20453758314743556
%100 0.16945660745403598
%500 0.05673948926098376
%1000 0.01746077346868398
%2000 0.0034813753078129675
%};
%\addlegendentry{(b1), $\tilde{\Gamma}_\infty$}
%\addplot [thick, red, mark=o, mark size=3, mark options={solid}]
%table {%
%10 0.23243340695265863
%50 0.22679989054814345
%100 0.227005023908476
%500 0.15207852111320974
%1000 0.09190268994786673
%200 0.03259540243747972
%};
%\addlegendentry{(a1), $\tilde{\Gamma}_\infty$}
%\addplot [semithick, color1, mark=*, mark %size=3, mark options={solid}]
%table {%
%10 9.61123966722385
%50 19.6098558288111
%100 19.0578485443594
%500 3.84840096852308
%1000 0.914896480934001
%2000 0.163167222343652
%};
%\addlegendentry{(b1), $\tilde{\Gamma}_1$}
%\addplot [semithick, green!50.0!black, %mark=*, mark size=3, mark options={solid}]
%table {%
%10 10.1943203764577
%50 24.1305538348769
%100 29.4936002734538
%500 24.6168102935663
%1000 12.2820186324181
%2000 2.08442780152757
%};
%\addlegendentry{(a1), $\tilde{\Gamma}_1$}
\addplot [thick, blue, mark=square, mark size=3, mark options={solid}]
table {%
10 0
50 0
100 0
500 0
1000 0
2000 0
};
\addlegendentry{Strong L. Lap.}
\addplot [thick, red, mark=square, mark size=3, mark options={solid}]
table {%
10 0.006266436085658513
50 0
100 0
500 0
1000 0
2000 0
};
\addlegendentry{Strong L. Adj.}
%\addplot [thick, black, mark=square, mark size=3, mark options={solid}]
%table {%
%10 0
%50 0
%100 0
%500 0
%1000 0
%2000 0
%};
%\addlegendentry{(b3), $\tilde{\Gamma}_\infty$}
%\addplot [thick, red, mark=square, mark size=3, mark options={solid}]
%table {%
%10 0.006122012812980654
%50 0
%100 0
%500 0
%1000 0
52000 0
%};
%\addlegendentry{(a3), $\tilde{\Gamma}_\infty$}
%\addplot [semithick, color1, %mark=triangle*, mark size=3, mark %options={solid,rotate=180}]
%table {%
%10 0
%50 0
%100 0
%500 0
%1000 0
%2000 0
%};
%\addlegendentry{(b3), $\tilde{\Gamma}_1$}
%\addplot [semithick, green!50.0!black, mark=triangle*, mark size=3, mark options={solid,rotate=180}]
%table {%
%10 0.0231918601266814
%50 0
%100 0
%500 0
%1000 0
%2000 0
%};
%\addlegendentry{(a3), $\tilde{\Gamma}_1$}
\end{axis}

\end{tikzpicture}}
\resizebox{.47\linewidth}{!}{% This file was created by tikzplotlib v0.8.2.
\begin{tikzpicture}

\definecolor{color0}{rgb}{0,0.75,0.75}
\definecolor{color1}{rgb}{0.75,0.75,0}

\begin{axis}[
legend cell align={left},
legend columns=2,
legend style={fill opacity=0.8, draw opacity=1, text opacity=1,at={(0.5,0.01)}, anchor=south, draw=white!80.0!black},
log basis x={10},
log basis y={10},
tick align=outside,
tick pos=both,
title={},
x grid style={white!69.01960784313725!black},
xlabel={\large Sample size $m$},
xmajorgrids,
xmin=7.67270499010925, xmax=2606.64264112612,
xmode=log,
xtick style={color=black},
y grid style={white!69.01960784313725!black},
ylabel={},
ymajorgrids,
ymin=0.0, ymax=0.8,
%ymode=log,
ytick style={color=black},
width=7.5cm,height=4cm,
grid=both,grid style={line width=.2pt, draw=gray!50},
]
\addplot [thick, blue, mark=o, mark size=3, mark options={solid}]
table {%
10 0.6761955517656386
50 0.6828097430899372
100 0.67936265068853
500 0.6791880025253135
1000 0.6840866551681152
2000 0.6869773138865607
};
\addlegendentry{Weak H. Lap.}
\addplot [thick, red, mark=o, mark size=3, mark options={solid}]
table {%
10 0.6723223486350515
50 0.6792020627014542
100 0.675626314295293
500 0.6721719991804816
1000 0.6672779399937779
2000 0.65129870506654
};
\addlegendentry{Weak H. Adj.}
%\addplot [thick, black, mark=o, mark size=3, mark options={solid}]
%table {%
%10 0.2312342855887354
%50 0.22842509790020504
%100 0.240025142630261
%500 0.22981089297050647
%1000 0.22961197913232423
%2000 0.2308653433600404
%};
%\addlegendentry{(b2), $\tilde{\Gamma}_\infty$}
%\addplot [thick, red, mark=o, mark size=3, mark options={solid}]
%table {%
%10 0.23218608289142095
%50 0.23147092448534398
%100 0.24394026357181203
%500 0.25142931726753814
%1000 0.26508612914754104
%2000 0.2802917400575734
%};
%\addlegendentry{(a2), $\tilde{\Gamma}_\infty$}
%\addplot [semithick, color1, mark=*, mark %size=3, mark options={solid}]
%table {%
%10 11.6702194246102
%50 27.3381285965847
%100 37.8163237941643
%500 85.6938891385243
%1000 122.566022596983
%2000 174.941763798429
%};
%\addlegendentry{(b2), $\tilde{\Gamma}_1$}
%\addplot [semithick, green!50.0!black, %mark=*, mark size=3, mark options={solid}]
%table {%
%10 11.5408725093683
%50 27.0854413920876
%100 37.1503786228246
%500 81.93939723853
%1000 113.881099304701
%2000 154.123267707622
%};
%\addlegendentry{(a2), $\tilde{\Gamma}_1$}
\addplot [thick, blue, mark=square, mark size=3, mark options={solid}]
table {%
10 0.6954544821578729
50 0.6960859449943019
100 0.69601286332849
500 0.6971300600725616
1000 0.6966681690488258
2000 0.6968399585781987
};
\addlegendentry{Strong H. Lap.}
\addplot [thick, red, mark=square, mark size=3, mark options={solid}]
table {%
10 0.6005593956081415
50 0.5307315605654379
100 0.521632287067596
500 0.5160098568709585
1000 0.5242164275682429
2000 0.522600037474196
};
\addlegendentry{Strong H. Adj.}
%\addplot [thick, black, mark=square, mark size=3, mark options={solid}]
%table {%
%10 0.23833540049283694
%50 0.23032292199027124
%100 0.23331721006023498
%500 0.23368292751140574
%1000 0.2325313449853039
%2000 0.23265863941479342
%};
%\addlegendentry{(b4), $\tilde{\Gamma}_\infty$}
%\addplot [thick, red, mark=square, mark size=3, mark options={solid}]
%table {%
%10 0.3495865122689027
%50 0.29707344138007874
%100 0.289305388597131
%500 0.2782719816776364
%1000 0.28566468726551386
%2000 0.2852534323515082
%};
%\addlegendentry{(a4), $\tilde{\Gamma}_\infty$}
%\addplot [semithick, color1, %mark=triangle*, mark size=3, mark %options={solid,rotate=180}]
%table {%
%10 12.4546070691433
%50 27.9225031134401
%100 39.4128812297225
%500 88.0684985895953
%1000 124.631300021197
%2000 176.301106589174
%};
%\addlegendentry{(b4), $\tilde{\Gamma}_1$}
%\addplot [semithick, green!50.0!black, %mark=triangle*, mark size=3, mark %options={solid,rotate=180}]
%table {%
%10 8.71415035895489
%50 17.4946462410798
%100 24.3696147239561
%500 54.0426694665729
%1000 77.3646622737918
%2000 109.142302769563
%};
%\addlegendentry{(a4), $\tilde{\Gamma}_1$}
\end{axis}

\end{tikzpicture}}
\resizebox{.52\linewidth}{!}{% This file was created by tikzplotlib v0.8.2.
\begin{tikzpicture}

\definecolor{color0}{rgb}{0,0.75,0.75}

\begin{axis}[
legend cell align={left},
legend style={at={(0.03,0.97)}, anchor=north west, draw=white!80.0!black},
log basis x={10},
tick align=outside,
tick pos=both,
title={},
x grid style={white!69.01960784313725!black},
xlabel={\large Graph size $n$},
xmajorgrids,
xmin=8.22340159426889, xmax=608.020895329328,
xmode=log,
xtick style={color=black},
y grid style={white!69.01960784313725!black},
ylabel={\large Scoring fct.~$\Gamma(\widehat{\bm v}_1)$},
ymajorgrids,
ymin=-0.0306493980923131, ymax=0.643637359938576,
ytick style={color=black},
width=7.5cm,height=4cm,
grid=both,grid style={line width=.2pt, draw=gray!50},
]
\addplot [thick, blue, mark=o, mark size=3, mark options={solid}]
table {%
10 8.68581233304996e-06
20 2.28976528512918e-06
50 0.00155007403721116
100 0.0159719960206891
200 0.0911942022285519
500 0.29801915507313
};
\addlegendentry{Weak L. Lap.}
\addplot [thick, red, mark=o, mark size=3, mark options={solid}]
table {%
10 0.00034005662322019
20 0.000618170662397516
50 0.0302311491462239
100 0.109300132900631
200 0.343319063544963
500 0.612987961846263
};
\addlegendentry{Weak L. Adj.}
%\addplot [thick, black, mark=o, mark size=3, mark options={solid}]
%table {%
%10 8.68581233304996e-06
%20 2.28976528512918e-06
%50 0.00152850127039684
%100 0.0141251615609872
%200 0.052798332964393
%500 0.0846893173816972
%};
%\addlegendentry{(b1), $\tilde{\Gamma}_\infty$}
%\addplot [thick, red, mark=o, mark size=3, mark options={solid}]
%table {%
%10 0.00034005662322019
%20 0.000618170662397516
%50 0.0277262308013456
%100 0.0752828230038913
%200 0.138391702821479
%500 0.133448471703925
%};
%\addlegendentry{(a1), $\tilde{\Gamma}_\infty$}
\addplot [thick, blue, mark=square, mark size=3, mark options={solid}]
table {%
10 2.54461424044179e-10
20 0
50 0
100 0
200 0
500 0
};
\addlegendentry{Strong L. Lap.}
\addplot [thick, red, mark=square, mark size=3, mark options={solid}]
table {%
10 8.96491860040222e-05
20 0
50 0
100 0
200 0
500 0
};
\addlegendentry{Strong L. Adj.}
%\addplot [thick, black, mark=square, mark size=3, mark options={solid}]
%table {%
%10 2.54461424044179e-10
%20 0
%50 0
%100 0
%200 0
%500 0
%};
%\addlegendentry{(b3), $\tilde{\Gamma}_\infty$}
%\addplot [thick, red, mark=square, mark size=3, mark options={solid}]
%table {%
%10 3.26015307030559e-05
%20 0
%50 0
%100 0
%200 0
%500 0
%};
%\addlegendentry{(a3), $\tilde{\Gamma}_\infty$}
\end{axis}

\end{tikzpicture}}
\resizebox{.47\linewidth}{!}{% This file was created by tikzplotlib v0.8.2.
\begin{tikzpicture}

\definecolor{color0}{rgb}{0,0.75,0.75}

\begin{axis}[
legend cell align={left},
legend columns=2,
legend style={at={(0.5,0.02)}, anchor=south, draw=white!80.0!black},
log basis x={10},
tick align=outside,
tick pos=both,
title={},
x grid style={white!69.01960784313725!black},
xlabel={\large Graph size $n$},
xmajorgrids,
xmin=8.22340159426889, xmax=608.020895329328,
xmode=log,
xtick style={color=black},
y grid style={white!69.01960784313725!black},
ylabel={},
ymajorgrids,
ymin=0, ymax=0.8,
ytick style={color=black},
width=7.5cm,height=4cm,
grid=both,grid style={line width=.2pt, draw=gray!50},
]
\addplot [thick, blue, mark=o, mark size=3, mark options={solid}]
table {%
10 0.656017708631502
20 0.669110531809037
50 0.679101773916012
100 0.68446501438616
200 0.69048639381957
500 0.69619593502667
};
\addlegendentry{Weak H. Lap.}
\addplot [thick, red, mark=o, mark size=3, mark options={solid}]
table {%
10 0.599489689641381
20 0.62102274404129
50 0.645824746069428
100 0.669331638898588
200 0.683837577563627
500 0.694458858475628
};
\addlegendentry{Weak H. Adj.}
%\addplot [thick, black, mark=o, mark size=3, mark options={solid}]
%table {%
%10 0.429304452591907
%20 0.366274220018179
%50 0.285192430391183
%100 0.229075294658722
%200 0.178613575598467
%500 0.126642645168041
%};
%\addlegendentry{(b2), $\tilde{\Gamma}_\infty$}
%\addplot [thick, red, mark=o, mark size=3, mark options={solid}]
%table {%
%10 0.412412664867149
%20 0.37835568595041
%50 0.318734008269456
%100 0.264454767955463
%200 0.20330894579288
%500 0.138371981165158
%};
%\addlegendentry{(a2), $\tilde{\Gamma}_\infty$}
\addplot [thick, blue, mark=square, mark size=3, mark options={solid,rotate=180}]
table {%
10 0.666013994081783
20 0.682311991887698
50 0.692155428092084
100 0.696828850448958
200 0.700283297410448
500 0.702285919101751
};
\addlegendentry{Strong H. Lap.}
\addplot [thick, red, mark=square, mark size=3, mark options={solid}]
table {%
10 0.504745969816761
20 0.53825775655468
50 0.534042949864178
100 0.519885381450608
200 0.512072594203776
500 0.474064387451538
};
\addlegendentry{Strong H. Adj.}
%\addplot [thick, black, mark=square, mark size=3, mark options={solid}]
%table {%
%10 0.41260018598675
%20 0.346571231882171
%50 0.277904843002223
%100 0.23187743328465
%200 0.191890321235032
%500 0.145816738857195
%};
%\addlegendentry{(b4), $\tilde{\Gamma}_\infty$}
%\addplot [thick, red, mark=square, mark size=3, mark options={solid}]
%table {%
%10 0.298211500734297
%20 0.328813047710142
%50 0.303019699325878
%100 0.278548610459078
%200 0.269893474551486
%500 0.23221939400601
%};
%\addlegendentry{(a4), $\tilde{\Gamma}_\infty$}
\end{axis}

\end{tikzpicture}}\vspace{-.3cm}
\caption{Effects of $m,n$ on the scoring function $\Gamma(\widehat{\bm v}_1)$: (Left) lowpass filter (${\cal T}_0$) and (Right) high pass filter (${\cal T}_1$). L./H. means low/highpass and Lap./Adj. means Laplacian /Adjacency.}\label{fig:syn_data_gammavsm}\vspace{-.4cm}
\end{figure}
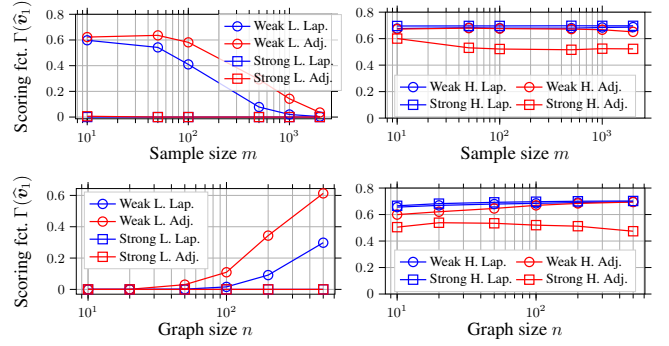

% \begin{figure}[t]
% \centering

% \caption{$\Gamma_2$, $\Gamma_\infty$ against the network size $n$ with (Left) low pass filter and (Right) high pass filter.}\label{fig:syn_data_gammavsn}\vspace{-.2cm}
% \end{figure}

\begin{figure*}[t]
\centering
\resizebox{.24\linewidth}{!}{% This file was created by tikzplotlib v0.8.2.
\begin{tikzpicture}

\definecolor{color0}{rgb}{0,0.75,0.75}

\begin{axis}[
legend cell align={left},
legend style={font = \large, fill opacity=0.8, draw opacity=1, text opacity=1,at={(0.03,0.97)}, anchor=north west, draw=white!80.0!black},
log basis x={10},
tick align=outside,
tick pos=both,
title={},
x grid style={white!69.01960784313725!black},
xlabel={\Large Graph size $n$},
xmajorgrids,
xmin=8.22340159426889, xmax=608.020895329328,
xmode=log,
xtick style={color=black},
y grid style={white!69.01960784313725!black},
ylabel={\Large Error rate},
ymajorgrids,
ymin=-0.0247, ymax=0.5187,
ytick style={color=black},
width=7cm,height=5.7cm,
grid=both,grid style={line width=.2pt, draw=gray!50},
]
\addplot [thick, blue, mark=o, mark size=3, mark options={solid}]
table {%
10 0
20 0
50 0
100 0
200 0
500 0.018
};
\addlegendentry{Weak Lap. }
\addplot [thick, red, mark=square, mark size=3, mark options={solid}]
table {%
10 0.000499999999999945
20 0
50 0.000499999999999945
100 0.005
200 0.094
500 0.371
};
\addlegendentry{Weak Adj. }

\addplot [thick, green!50!black, mark=*, mark size=3, mark options={solid}]
table {%
10 0
20 0
50 0
100 0
200 0
500 0
};
\addlegendentry{Strong Lap. }
\addplot [thick, black, mark=diamond, mark size=5, mark options={solid}]
table {%
10 0.000499999999999945
20 0
50 0
100 0
200 0
500 0
};
\addlegendentry{Strong Adj. }
\addplot [dashed,thick,green!50!black, mark=*, mark size=3, mark options={solid}]
table {%
10 0
20 0
50 0
100 0
200 0
500 0
};
%\addlegendentry{(b3)+(b4),$\Gamma_\infty$ }
\addplot [dashed,thick, black, mark=diamond, mark size=5, mark options={solid}]
table {%
10 0.000499999999999945
20 0
50 0
100 0
200 0
500 0
};
%\addlegendentry{(a3)+(a4),$\Gamma_\infty$ }
\addplot [dashed,thick, blue, mark=o, mark size=3, mark options={solid}]
table {%
10 0
20 0
50 0
100 0
200 0.001
500 0.1065
};
%\addlegendentry{(b1)+(b2),$\Gamma_\infty$ }
\addplot [dashed,thick, red, mark=square, mark size=3, mark options={solid}]
table {%
10 0.000499999999999945
20 0
50 0.003
100 0.016
200 0.2335
500 0.494
};
%\addlegendentry{(a1)+(a2),$\Gamma_\infty$ }
\end{axis}

\end{tikzpicture}}
\resizebox{.24\linewidth}{!}{% This file was created by tikzplotlib v0.8.2.
\begin{tikzpicture}

\definecolor{color0}{rgb}{0,0.75,0.75}
\definecolor{color1}{rgb}{0.75,0.75,0}

\begin{axis}[
legend cell align={left},
legend columns=3,
legend style={fill opacity=0.8, draw opacity=1, text opacity=1,at={(0.5,1.4)}, anchor=north, draw=white!80.0!black},
log basis x={10},
tick align=outside,
tick pos=both,
title={},
x grid style={white!69.01960784313725!black},
xlabel={\large Sample size $m$},
xmajorgrids,
xmin=7.67270499010925, xmax=2606.64264112612,
xmode=log,
xtick style={color=black},
y grid style={white!69.01960784313725!black},
ylabel={},
ymajorgrids,
ymin=-0.0251, ymax=0.5271,
ytick style={color=black},
width=6.5cm,height=4.9cm,
grid=both,grid style={line width=.2pt, draw=gray!50},
]
\addplot [thick, blue, mark=o, mark size=3, mark options={solid}]
table {%
10 0.4325
50 0.46
100 0.1925
500 0
1000 0
2000 0
};
%\addlegendentry{(b1)+(b2),$\tilde{\Gamma}_2$ }
\addplot [thick, red, mark=square, mark size=3, mark options={solid}]
table {%
10 0.4725
50 0.4835
100 0.374
500 0.083
1000 0.00350000000000006
2000 0
};
%\addlegendentry{(a1)+(a2),$\tilde{\Gamma}_2$ }
\addplot [dashed,thick, blue, mark=o, mark size=3, mark options={solid}]
table {%
10 0.502
50 0.4845
100 0.3055
500 0
1000 0
2000 0
};
%\addlegendentry{(b1)+(b2),$\tilde{\Gamma}_\infty$ }
\addplot [dashed,thick, red, mark=square, mark size=3, mark options={solid}]
table {%
10 0.5
50 0.499
100 0.446
500 0.181
1000 0.0395
2000 0.003
};
%\addlegendentry{(a1)+(a2),$\tilde{\Gamma}_\infty$ }
%\addplot [semithick, color1, mark=*, mark %size=3, mark options={solid}]
%table {%
%10 0.4125
%50 0.4715
%100 0.1895
%500 0
%1000 0
%2000 0
%};
%\addlegendentry{H(A), l1 norm}
%\addplot [semithick, green!50.0!black, %mark=*, mark size=3, mark options={solid}]
%table {%
%10 0.4515
%50 0.489
%100 0.3705
%500 0.0755
%1000 0.002
%2000 0
%};
%\addlegendentry{H(L), l1 norm}
\addplot [thick, green!50!black, mark=*, mark size=3, mark options={solid}]
table {%
10 0
50 0
100 0
500 0
1000 0
2000 0
};
%\addlegendentry{(b3)+(b4),$\tilde{\Gamma}_2$ }
\addplot [thick, black, mark=diamond, mark size=5, mark options={solid}]
table {%
10 0.052
50 0.148
100 0
500 0
1000 0
2000 0
};
%\addlegendentry{(a3)+(a4),$\tilde{\Gamma}_2$ }
\addplot [dashed,thick, green!50!black, mark=*, mark size=3, mark options={solid}]
table {%
10 0
50 0
100 0
500 0
1000 0
2000 0
};
%\addlegendentry{(b3)+(b4),$\tilde{\Gamma}_\infty$ }
\addplot [dashed,thick, black, mark=diamond, mark size=5, mark options={solid}]
table {%
10 0.0175000000000001
50 0.107
100 0.0125
500 0
1000 0
2000 0
};
%\addlegendentry{(a3)+(a4),$\tilde{\Gamma}_\infty$ }
%\addplot [semithick, color1, %mark=triangle*, mark size=3, mark %options={solid,rotate=180}]
%table {%
%10 0
%50 0
%100 0
%500 0
%1000 0
%2000 0
%};
%\addlegendentry{strong H(A), l1 norm}
%\addplot [semithick, green!50.0!black, %mark=triangle*, mark size=3, mark %options={solid,rotate=180}]
%table {%
%10 0.0580000000000001
%50 0.1195
%100 0
%500 0
%1000 0
%2000 0
%};
%\addlegendentry{strong H(L), l1 norm}
\end{axis}

\end{tikzpicture}}
\resizebox{.24\linewidth}{!}{% This file was created by tikzplotlib v0.8.2.
\begin{tikzpicture}

\definecolor{color0}{rgb}{0,0.75,0.75}

\begin{axis}[
legend cell align={left},
legend columns=4,
legend style={at={(0.5,1.2)}, anchor=north, draw=white!80.0!black},
log basis x={10},
tick align=outside,
tick pos=both,
title={},
x grid style={white!69.01960784313725!black},
xlabel={\large Noise variance $\sigma^2$},
xmajorgrids,
xmin=0.0891250938133746, xmax=1.12201845430196,
xmode=log,
xtick style={color=black},
y grid style={white!69.01960784313725!black},
ymajorgrids,
ymin=-0.0217, ymax=0.4557,
ytick style={color=black},
width=6.5cm,height=4.9cm,
x dir = reverse,
grid=both,grid style={line width=.2pt, draw=gray!50},
]
\addplot [thick, blue, mark=o, mark size=3, mark options={solid}]
table {%
0.01 0
%0.04 0
0.09 0
%0.16 0
0.25 0
%0.36 0
0.49 0
%0.64 0
0.81 0
1 0.001
};
%\addlegendentry{(b1)+(b2),$\tilde{\Gamma}_2$ }
\addplot [thick, red, mark=square, mark size=3, mark options={solid}]
table {%
0.01 0.001
%0.04 0.00700000000000001
0.09 0.00700000000000001
%0.16 0.021
0.25 0.031
%0.36 0.0555
0.49 0.122
%0.64 0.201
0.81 0.2435
1 0.3075
};
%\addlegendentry{(a1)+(a2),$\tilde{\Gamma}_2$ }
\addplot [dashed,thick, blue, mark=o, mark size=3, mark options={solid}]
table {%
0.01 0
%0.04 0
0.09 0
%0.16 0
0.25 0.000499999999999945
%0.36 0.000499999999999945
0.49 0.000499999999999945
%0.64 0.00150000000000006
0.81 0.00649999999999995
1 0.013
};
%\addlegendentry{(b1)+(b2),$\tilde{\Gamma}_\infty$ }
\addplot [dashed,thick, red, mark=square, mark size=3, mark options={solid}]
table {%
0.01 0.0195000000000001
%0.04 0.036
0.09 0.046
%0.16 0.0755
0.25 0.113
%0.36 0.17
0.49 0.2435
%0.64 0.3415
0.81 0.3885
1 0.434
};
%\addlegendentry{(a1)+(a2),$\tilde{\Gamma}_\infty$ }
\addplot [thick, green!50!black, mark=*, mark size=3, mark options={solid}]
table {%
0.01 0
%0.04 0
0.09 0
%0.16 0
0.25 0
%0.36 0
0.49 0
%0.64 0
0.81 0
1 0
};
%\addlegendentry{(b3)+(b4),$\tilde{\Gamma}_2$ }
\addplot [thick, black, mark=diamond, mark size=5, mark options={solid}]
table {%
0.01 0
%0.04 0
0.09 0
%0.16 0
0.25 0
%0.36 0
0.49 0
%0.64 0
0.81 0
1 0
};
%\addlegendentry{(a3)+(a4),$\tilde{\Gamma}_2$ }
\addplot [dashed,thick, green!50!black, mark=*, mark size=3, mark options={solid}]
table {%
0.01 0
%0.04 0
0.09 0
%0.16 0
0.25 0
%0.36 0
0.49 0
%0.64 0
0.81 0.000499999999999945
1 0.001
};
%\addlegendentry{(b3)+(b4),$\tilde{\Gamma}_\infty$ }
\addplot [dashed,thick, black, mark=diamond, mark size=5, mark options={solid}]
table {%
0.01 0
%0.04 0
0.09 0
%0.16 0
0.25 0
%0.36 0
0.49 0
%0.64 0
0.81 0
1 0
};
%\addlegendentry{(a3)+(a4),$\tilde{\Gamma}_\infty$ }
\end{axis}

\end{tikzpicture}}
\resizebox{.24\linewidth}{!}{% This file was created by tikzplotlib v0.9.4.
\begin{tikzpicture}

\begin{axis}[
legend cell align={left},
legend columns=3,
legend style={fill opacity=0.8, draw opacity=1, text opacity=1, at={(0.97,0.03)}, anchor=south east, draw=white!80!black},
tick align=outside,
tick pos=left,
title={},
x grid style={white!69.0196078431373!black},
xlabel={\large Graph frequency $i$},
xmin=1, xmax=103.9,
xtick style={color=black},
xmode = log,
y grid style={white!69.0196078431373!black},
ylabel={\large Scoring fct.},
ymin=-0.05, ymax=0.8,
ytick style={color=black},
width=6.25cm,height=5.2cm,
grid=both,grid style={line width=.2pt, draw=gray!50},
xtick = {1,2,10,20,99},
xticklabels = {1,2,10,20,99}
]
\addplot [thick, blue]
table {%
1 0
2 0.688045095186156
3 0.700873538629553
4 0.671635702306291
5 0.583643400793708
6 0.51390088185826
7 0.689348352884022
8 0.497419497592912
9 0.671163017904645
10 0.617928784370482
11 0.687145186495596
12 0.676187896029628
13 0.66203152228894
14 0.622394105121564
15 0.696754836634768
16 0.678192930055623
17 0.674873842748447
18 0.667261110860049
19 0.67955665488132
20 0.680320618924369
21 0.650881017265632
22 0.693594617022071
23 0.688573843167782
24 0.683902201698102
25 0.683783868499222
26 0.703294503940398
27 0.69345014468726
28 0.684748797095972
29 0.690752186781216
30 0.669646128488753
31 0.67982336346792
32 0.68513519708432
33 0.692423066229832
34 0.675690331034539
35 0.654185282021735
36 0.664324147646174
37 0.695144736093621
38 0.689025883047666
39 0.636050944209173
40 0.693226186016957
41 0.706033979211118
42 0.64603674059492
43 0.703810379062893
44 0.689055233923504
45 0.673947528255531
46 0.688432335583347
47 0.676066807192391
48 0.703154822685149
49 0.652431012839582
50 0.674131137161866
51 0.668186069864758
52 0.677650994013602
53 0.677508628292564
54 0.665073296495496
55 0.699972669312874
56 0.691156364939297
57 0.676946864972485
58 0.705532271411959
59 0.698368039506027
60 0.703018882331369
61 0.693072447071052
62 0.685957473805972
63 0.643719487129157
64 0.648727071161539
65 0.700614315596807
66 0.697625242554826
67 0.671661845087274
68 0.686120182075472
69 0.693802487688646
70 0.683022465957705
71 0.701287657524957
72 0.652901141729909
73 0.665035158632706
74 0.692679132163468
75 0.643822095383864
76 0.705660004309182
77 0.669296531178981
78 0.706445938184262
79 0.623364852816729
80 0.690110785442967
81 0.702377965073948
82 0.68608692554728
83 0.682641852369233
84 0.704520280098735
85 0.698195786279807
86 0.697460349932251
87 0.665583201002465
88 0.674139167439949
89 0.706080644147155
90 0.689686249443882
91 0.686622845841561
92 0.70443309876439
93 0.695972074725057
94 0.67822862133431
95 0.631299622088672
96 0.663327183204057
97 0.685753130616425
98 0.667301326175968
99 0.672741604607294
};
\addlegendentry{$\Gamma(\widehat{\bm v}_i)$}
\addplot [thick, red]
table {%
1 0
2 0.180775649852099
3 0.173309384779329
4 0.20955254600536
5 0.179127937849587
6 0.14206872448051
7 0.36671065928346
8 0.20894629605406
9 0.224991903561787
10 0.228338835087119
11 0.270102911716225
12 0.283494035046924
13 0.243057608139786
14 0.252678178060174
15 0.293102094033051
16 0.2782511114918
17 0.266796345565491
18 0.212135293191641
19 0.26432454538796
20 0.246498164866602
21 0.265778522124974
22 0.375452688438314
23 0.31117244294021
24 0.233654908730923
25 0.240521999563154
26 0.272249845017623
27 0.33034369656365
28 0.272825166483187
29 0.270179649726383
30 0.232577836922723
31 0.261150849402338
32 0.217063564357458
33 0.245208988538673
34 0.219707440727622
35 0.223237729865792
36 0.178510357544762
37 0.274841673292923
38 0.255326327731361
39 0.211106326591904
40 0.30046787414516
41 0.267042434232188
42 0.1929402068884
43 0.696829538816338
44 0.200928611685159
45 0.257267969206808
46 0.282706084134052
47 0.232588885863995
48 0.231004824743214
49 0.244194073828885
50 0.224343037297866
51 0.204393003194845
52 0.287002773519047
53 0.42030293962051
54 0.218176755963224
55 0.252279846977759
56 0.244647660754987
57 0.204198232739079
58 0.407365188755485
59 0.224776188455383
60 0.3521381909224
61 0.278147996948769
62 0.221405339379103
63 0.210464290467693
64 0.26378700705406
65 0.231155497203591
66 0.281332674944184
67 0.24071096417325
68 0.265238678558695
69 0.300716658343186
70 0.255846516101525
71 0.279457371969514
72 0.283935030635511
73 0.261993563855918
74 0.316185479797716
75 0.283530332504576
76 0.390700773369378
77 0.225424294815254
78 0.21814394125484
79 0.238131151193853
80 0.256951484922304
81 0.248779582461725
82 0.345332553649215
83 0.276796373384218
84 0.243118846340678
85 0.210115117593798
86 0.293695847967195
87 0.234922417799836
88 0.261238675659842
89 0.226638027871669
90 0.233664136381102
91 0.297538502080513
92 0.319199833066734
93 0.208827965849294
94 0.199288129747407
95 0.272384513898555
96 0.232149290702771
97 0.213566919237627
98 0.241408151369539
99 0.211472254027643
};
\addlegendentry{$\Gamma(\widehat{\bm v}_i)_\infty$}
\end{axis}

\end{tikzpicture}}\vspace{-0.4cm}
\caption{Error rate against (Left) the number of samples $m$, (Middle-left) the graph size $n$, and (Middle-right) the noise variance $\sigma^2$. Lap./Adj. means Laplacian/Adjacency. Dashed lines are the detector with scoring function $\Gamma({\bm v})_\infty \eqdef \| {\bm v} - ({\bm v} )_+ \|_\infty \wedge \| {\bm v} + (-{\bm v} )_+ \} \|_\infty$ with the same respective color/marker. (Right) Scoring function $\Gamma(\widehat{\bm v}_i)$ and $\Gamma(\widehat{\bm v}_i)_\infty$ against the graph frequency of the {\sf Stock} data's sample covariance.}\label{fig:syn_data_errorrate}\vspace{-.4cm}
\end{figure*}
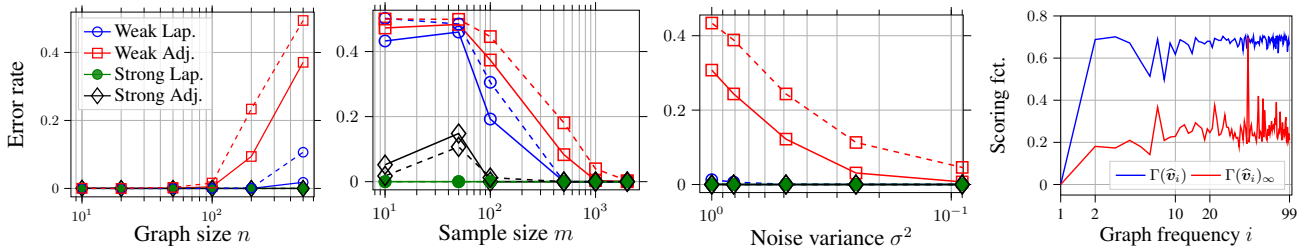

\section{Numerical Experiments}\vspace{-.1cm} \label{sec:num}
% This section performs numerical experiments for the proposed detector on synthetic and real data.\vspace{.1cm}

\noindent\textbf{Synthetic Data.} For the experiments below, the $n$-node graph $G$ is generated as an undirected Erdos-Renyi (ER) graph with connection probability of $p = 2\log(n)/n$. Each observed graph signal ${\bm y}_\ell \in \Re^{n}$ is independently generated as ${\bm y}_\ell = {\cal H}({\bm S}){\bm x}_\ell+ {\bm w}_\ell$ with excitation ${\bm x}_\ell \sim N({\bm 0},{\bm I})$ and noise ${\bm w}_\ell \sim N({\bm 0},\sigma^2{\bm I})$. ${\cal H}({\bm S})$ is a graph filter with two possible settings (a) ${\bm S} = {\bm L}$ is Laplacian matrix or (b) ${\bm S} = {\bm A}$ is binary adjacency matrix. 

We verify the analysis in Section~\ref{sec:det} by experimenting with four types of graph filters ${\cal H}({\bm S})$ and we take an example of highpass filter to represent the alternative hypothesis ${\cal T}_1$. The first set of experiments considers pairs of \emph{weak} lowpass/highpass filters as ${\cal H}({\bm S})$: \vspace{-0.2cm}
\beq\label{eq:weakpf}
\begin{split}
&\text{Setting (a)}\ \ \  ({\cal T}_0):\  ({\bm I}+\alpha {\bm L})^{-1},\  ({\cal T}_1):\  {\bm I}+\alpha {\bm L}, \\
&\text{Setting (b)}\ \ \ ({\cal T}_0):\ ({\bm I}-\alpha {\bm A})^{-1},\ ({\cal T}_1):\  {\bm I}-\alpha{\bm A}, \vspace{-0.1cm}
\end{split}
\eeq
where $\alpha = 0.5/d_{\max}$ with $d_{\max}$ being the highest degree of graph $G$.
The second set of experiments considers pairs of \emph{strong} lowpass/highpass filter as ${\cal H}({\bm S})$: let $\tau = 10/d_{max}$,\vspace{-0.1cm}
\beq\label{eq:strongpf}
\begin{split}
&\text{Setting (a)}\ \ \  ({\cal T}_0):\  e^{-\tau{\bm L}},\  ({\cal T}_1):\  e^{\tau{\bm L}},\\
&\text{Setting (b)}\ \ \ ({\cal T}_0):\ e^{\tau{\bm A}},\ ({\cal T}_1):\  e^{-\tau{\bm A}}.\\[-.1cm]
\end{split}
\eeq
It can be shown that the above filters under ${\cal T}_0$ are first-order lowpass with $\eta_1 \approx 1$ in \eqref{eq:weakpf}, and $\eta_1 \ll 1$ in \eqref{eq:strongpf}. 

% Recall that $\widehat{\bm v}_1$ is the top eigenvector of sample covariance matrix $\widehat{\bm C}_y^m$ \eqref{eq:samp_cov} and  $\Gamma(\widehat{\bm v}_1)$ is the scoring function in \eqref{eq:stat}. 
In Fig.~\ref{fig:syn_data_gammavsm}, we illustrate the effects of number of samples $m$ and graph size $n$ on $\Gamma(\widehat{\bm v}_1)$ [cf.~\eqref{eq:stat}]. Fixing the noise variance at $\sigma^2 = 0.01$, the averaged values of $\Gamma( \widehat{\bm v}_1 )$  over 1000 trials is plotted. 
In Fig.~\ref{fig:syn_data_gammavsm} (Top), we fix the graph size at $n=100$ and observe that under ${\cal T}_0$, the scoring function $\Gamma( \widehat{\bm v}_1 )$ decays as the number of samples $m$ grows. On the other hand, in Fig.~\ref{fig:syn_data_gammavsm} (Bottom), we fix the sample size at $m=1000$ and observe that under ${\cal T}_0$, $\Gamma( \widehat{\bm v}_1 )$ may increase as the graph size $n$ grows for weak lowpass filters. In both comparisons, under ${\cal T}_1$, the scoring function floats around a constant value $\geq 0.4$ irrespective of $m,n$.  
The above findings are consistent with \eqref{eq:gamma1}, \eqref{eq:gamma1_t1}, which predicts that under ${\cal T}_0$, the scoring function may increase as the graph size $n$ grows, thereby leading to the degraded detection performance with \eqref{eq:detector} as we shall illustrate next. 

We next examine the detection performance measured in terms of the error rate with equal number of data samples coming from pairs of weak and strong low/highpass filter in \eqref{eq:weakpf}, \eqref{eq:strongpf}. We define:\vspace{-.15cm}
\beq\label{eq:errorrate}
\text{\sf Error rate} \eqdef 0.5\cdot{\cal P}(\widehat{\cal T}={\cal T}_1|{\cal T}_0)+0.5\cdot{\cal P}(\widehat{\cal T}={\cal T}_0|{\cal T}_1),\vspace{-.1cm}
\eeq
and evaluate the averaged error rate from 1000 trials. 
For benchmarking purpose, we also simulate a similar detector as \eqref{eq:detector} but replace the scoring function in \eqref{eq:stat} with one that is computed by the $\ell_\infty$ norm, e.g., $\Gamma({\bm v})_\infty \eqdef \| {\bm v} - ({\bm v} )_+ \|_\infty \wedge \| {\bm v} + (-{\bm v} )_+ \} \|_\infty$.

The results are presented in Fig.~\ref{fig:syn_data_errorrate}. We compare the averaged error rate against graph size $n$, sample size $m$, noise variance $\sigma^2$, while fixing $(m,\sigma^2) = (1000,0.01)$, $(n,\sigma^2) = (100,0.01)$, $(n,m) = (100,1000)$, respectively.
For each parameter setting in the experiments, we generate $m$ samples for each hypothesis, i.e., with lowpass/highpass filter, in order to evaluate \eqref{eq:errorrate}. 
To distinguish between weak lowpass/highpass filter [cf.~\eqref{eq:weakpf}], a larger $m$ reduces the error rate while a larger $n$ raises the error rate. Moreover, the error rate reduces with a small noise variance. For the experiments with pairs of strong lowpass/highpass filter [cf.~\eqref{eq:strongpf}], the detection performance is almost invariant with $n,m,\sigma^2$, i.e., the error rate is close to 0 for all cases.
Lastly, we observe that the $\ell_2$ norm scoring function has consistently outperformed its $\ell_\infty$ norm counterpart.
The above results are consistent with the prediction in Section~\ref{sec:det}.\vspace{.1cm} 

\noindent \textbf{Real Data.}
We consider identifying first-order lowpass signals from 3 real datasets.  
The first dataset ({\sf Stock}) is the daily return from S\&P100 stocks in May 2018 to Aug 2019 with $n=99$ stocks, $m=300$ samples, collected from \url{https://www.alphavantage.co/}. The second dataset ({\sf Senate}) contains $m = 696$ votes grouped by $n=50$ states at the US Senate in 2007 to 2009, collected from \url{https://voteview.com}. The third dataset ({\sf COVID-19}) is the daily increment of COVID-19 confirmed cases in the US from May 5th 2020 to Oct 15th 2020 with $n = 44$ states, $m = 164$ samples, collected from \url{https://covidtracking.com/}. 
% The abbreviations of states' names also come from it.

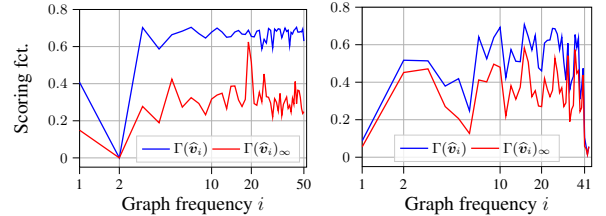
\begin{figure}[t]
\centering
\resizebox{.48\linewidth}{!}{% This file was created by tikzplotlib v0.9.4.
\begin{tikzpicture}

\begin{axis}[
legend cell align={left},
legend columns=3,
legend style={fill opacity=0.8, draw opacity=1, text opacity=1, at={(0.97,0.03)}, anchor=south east, draw=white!80!black},
tick align=outside,
tick pos=left,
title={},
x grid style={white!69.0196078431373!black},
xlabel={\large Graph frequency $i$},
xmin=1, xmax=52.45,
xtick style={color=black},
xmode = log,
y grid style={white!69.0196078431373!black},
ylabel={\large Scoring fct.},
ymin=-0.05, ymax=0.8,
ytick style={color=black},
width=6.5cm,height=5cm,
grid=both,grid style={line width=.2pt, draw=gray!50},
xtick = {1,2,10,20,50},
xticklabels = { 1, 2 ,10, 20,50}
]
\addplot [thick, blue]
table {%
1 0.407112602797137
2 0
3 0.702418882595241
4 0.586859643211537
5 0.663952893111599
6 0.683343216699121
7 0.703972754370855
8 0.670165726630657
9 0.645970356030845
10 0.677910171152001
11 0.699698550769962
12 0.678549948509865
13 0.650240689303688
14 0.653494354085524
15 0.67863587430945
16 0.686885031248283
17 0.667311864215963
18 0.665489100263036
19 0.688307184002318
20 0.654883254477419
21 0.65208099446683
22 0.683005446398227
23 0.687665716353049
24 0.587084964776147
25 0.639179719574882
26 0.619155540758083
27 0.69741035402551
28 0.687693844501451
29 0.591799079831272
30 0.694492319390646
31 0.683123224360004
32 0.696270679137331
33 0.693530561148091
34 0.64791632213312
35 0.673737270622972
36 0.699868663417523
37 0.681667203479191
38 0.688533953168028
39 0.604774885341022
40 0.676372343827313
41 0.667890942958713
42 0.686842584978182
43 0.694875544889023
44 0.704033194169483
45 0.678316634984987
46 0.67758879712498
47 0.681052673700568
48 0.67918791463625
49 0.694933053401414
50 0.629801805831614
};
\addlegendentry{$\Gamma(\widehat{\bm v}_i)$}
\addplot [thick, red]
table {%
1 0.149319964229075
2 0
3 0.277402481094638
4 0.189238441260853
5 0.424247511124306
6 0.273673417909828
7 0.324767667266621
8 0.293458472537297
9 0.232337468537429
10 0.317077764963535
11 0.338710809927414
12 0.348653968742924
13 0.265149974819543
14 0.386134573006004
15 0.387840268836691
16 0.311962359416505
17 0.277209086430158
18 0.315350485472725
19 0.624997167147666
20 0.515782411676509
21 0.292806172785015
22 0.283639236305132
23 0.353645104730997
24 0.302429871044347
25 0.452508837745915
26 0.329959632405179
27 0.326598412427487
28 0.295508735590342
29 0.22020981438054
30 0.319025992875981
31 0.300140717474045
32 0.252355293766968
33 0.347419243079568
34 0.228345329678488
35 0.288708566806943
36 0.309094088928464
37 0.327901133888571
38 0.26707826257212
39 0.283690806453581
40 0.247386320548717
41 0.320890250093061
42 0.289509779613398
43 0.411762907439774
44 0.291942771276019
45 0.364936825459896
46 0.361972511165366
47 0.325513481490325
48 0.2665817761911
49 0.239435392197385
50 0.249221010344361
};
\addlegendentry{$\Gamma(\widehat{\bm v}_i)_\infty$}
\end{axis}

\end{tikzpicture}}
\resizebox{.425\linewidth}{!}{% This file was created by tikzplotlib v0.9.4.
\begin{tikzpicture}

\begin{axis}[
legend cell align={left},
legend columns=3,
legend style={fill opacity=0.8, draw opacity=1, text opacity=1, at={(0.5,0.03)}, anchor=south, draw=white!80!black},
tick align=outside,
tick pos=left,
title={},
x grid style={white!69.0196078431373!black},
xlabel={\large Graph frequency $i$},
xmin=1, xmax=46.15,
xtick style={color=black},
xmode = log,
y grid style={white!69.0196078431373!black},
ylabel={},
ymin=-0.05, ymax=0.8,
ytick style={color=black},
width=6.5cm,height=5cm,
grid=both,grid style={line width=.2pt, draw=gray!50},
xtick = {1,2,10,20,41},
xticklabels = {1,2,10,20,41}
]
\addplot [thick, blue]
table {%
1 0.0867463265298839
2 0.517239277962707
3 0.513849809544016
4 0.379357528975549
5 0.41886239781245
6 0.247532752025083
7 0.641655618414086
8 0.524921301561663
9 0.638357827970671
10 0.692248931491177
11 0.424263591239695
12 0.573899146298889
13 0.511680938471076
14 0.509444458582625
15 0.707035865732309
16 0.64315649520688
17 0.57132636446778
18 0.649094541762451
19 0.47661923299552
20 0.460462885276353
21 0.608821036180143
22 0.623301535245572
23 0.625278121947511
24 0.686784530001095
25 0.685097334136126
26 0.656426293013223
27 0.573801353420094
28 0.641473936014444
29 0.504594258693944
30 0.410849029367821
31 0.608181866486056
32 0.360288795344322
33 0.405835109477837
34 0.431141274310609
35 0.653934071519775
36 0.584878549753338
37 0.634491473912895
38 0.309685619102825
39 0.370162061337466
40 0.473916124878866
41 0.104819821906073
42 0.0603524204551954
43 0.0167193970290611
44 0.0563769893017388
};
\addlegendentry{$\Gamma(\widehat{\bm v}_i)$}
\addplot [thick, red]
table {%
1 0.0551064544154846
2 0.451742356713423
3 0.471182593388206
4 0.270128465015594
5 0.206078297624735
6 0.126584674047617
7 0.412022725352861
8 0.401819189520451
9 0.496130341608782
10 0.479947053260866
11 0.219327321378721
12 0.381041365857524
13 0.309301848770361
14 0.373252742973832
15 0.581944573182101
16 0.508139363175497
17 0.322943478871078
18 0.354736085804527
19 0.273135703516803
20 0.269829198468969
21 0.419498124938079
22 0.36276798247475
23 0.32289159489605
24 0.356862354648224
25 0.526874238278545
26 0.454973358905001
27 0.343762685800397
28 0.370129532275459
29 0.270554036089615
30 0.290115424082657
31 0.541480264490014
32 0.189616437223819
33 0.254394292425663
34 0.276113528530881
35 0.573578337676237
36 0.447700917638539
37 0.461530031090107
38 0.244417188271147
39 0.26304681648213
40 0.438511584592936
41 0.0578855601881617
42 0.0461377913798131
43 0.0135008338400554
44 0.0562236877178647
};
\addlegendentry{$\Gamma(\widehat{\bm v}_i)_\infty$}
\end{axis}

\end{tikzpicture}}\vspace{-.3cm}
\caption{Scoring function $\Gamma(\widehat{\bm v}_i)$, $\Gamma(\widehat{\bm v}_i)_\infty$ against the graph frequency order. (Left) {\sf Senate} data (Right) {\sf COVID-19} data.}\label{fig:real_stocAvotedata}\vspace{-.4cm}
\end{figure}

\begin{figure}[t]
\centering
\includegraphics[height=.14\textheight]{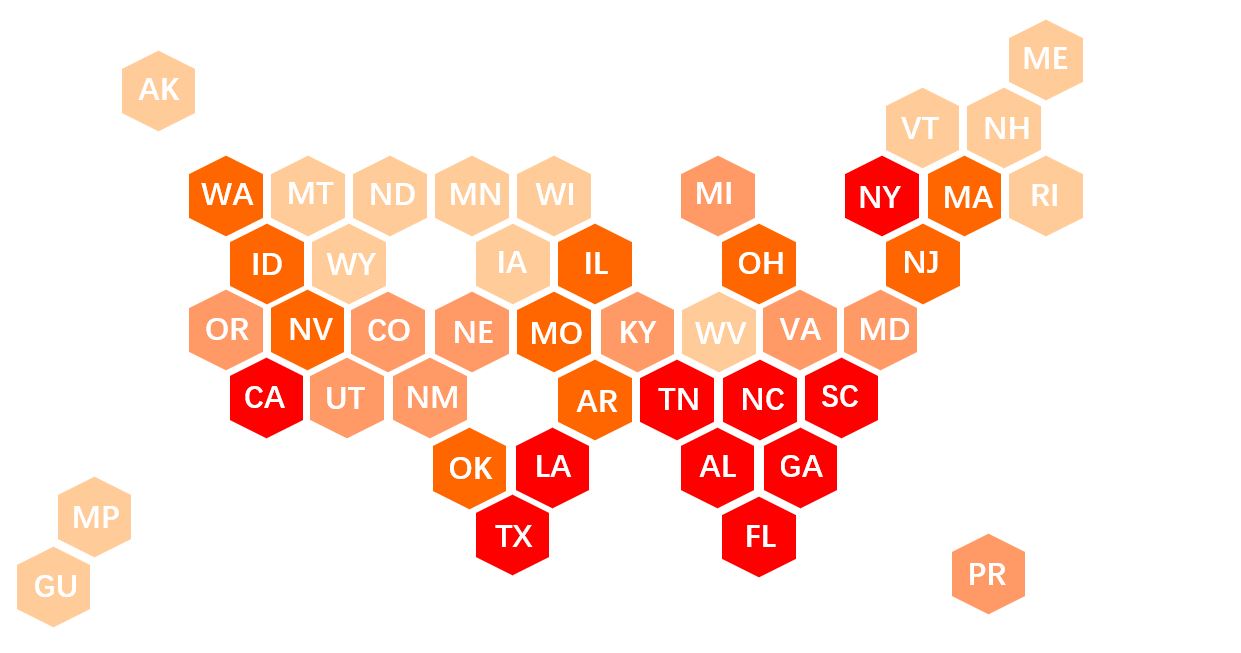}\vspace{-.3cm}
\caption{Estimated centrality of 44 states in {\sf COVID-19} data. Deeper color indicates the state has a higher centrality.}\label{fig:real_coviddata}\vspace{-.4cm}
\end{figure}

Fig.~\ref{fig:syn_data_errorrate} (Right), Fig.~\ref{fig:real_stocAvotedata} plots the scoring functions $\Gamma( \widehat{\bm v}_i )$ against the eigenvalue order $i$ for the 3 dataset considered. 
We first observe that the {\sf Stock} data would satisfy ${\cal T}_0$ since $\Gamma( \widehat{\bm v}_1 ) < \Gamma( \widehat{\bm v}_j )$ for all $j \neq 1$, indicating that it is likely to be a set of first-order lowpass graph signals. This certifies that lowpass GSP tools can be applied to the dataset. For example, applying the method from \cite{Wai2020BCest,Segarra2020BCest} ranks the centrality of stocks in the decreasing order as: NVDA, NFLX, AMZN, ADBE, PYPL, CAT, MA, GOOG, GOOGL, BA; see \cite{Wai2020BCest} for details. 
For the {\sf Senate} data, we have $\Gamma( \widehat{\bm v}_2 ) < \Gamma( \widehat{\bm v}_j )$ for all $j \neq 2$ which suggests that the data may not be first-order lowpass. However, since the minimum occur at $\Gamma( \widehat{\bm v}_2 )$, it is plausible that the data is generated from a lowpass graph filter with cutoff frequency at $\lambda_2$.
For the {\sf COVID-19} data, we see that $\Gamma( \widehat{\bm v}_1 ) < \Gamma( \widehat{\bm v}_j )$ for $2 \leq j \leq 40$, yet $\Gamma( \widehat{\bm v}_j)$ are small again for $j=41,...,44$. This suggests the sampled covariance has more than one (close-to) positive eigenvector. We suspect that this abnormally is due to outliers such as the beginning wave of a COVID-19 infection event at a state.

Inspired by the above, we model the {\sf COVID-19} data as a set of first-order lowpass graph signals (with outliers). Again, we can apply the blind centrality estimation method from \cite{Wai2020BCest,Segarra2020BCest} to rank the centrality of states. The results are illustrated in Fig.~\ref{fig:real_coviddata}. The top states ranked in decreasing centrality are FL, TX, CA, GA, LA, TN, SC, AL, NC, NY. Some of these states are the transportation hubs.\vspace{.1cm}

\noindent \textbf{Conclusions}. This paper utilizes the Perron-Frobenius theorem to design a simple, data-driven detector for identifying first-order lowpass graph signals. The detector can be used to provide certificates for applying lowpass GSP tools, and to make inference about the type of network dynamics. Future works include verifying \Cref{conj:hpf}, designing detectors for \emph{higher-order} lowpass graph signals.

\newpage
\bibliographystyle{IEEEtran}
\bibliography{ref_list}

\end{document}